\ifx\pdftexversion\undefined
  \documentclass[10pt,preprint,dvips]{aastex}
\else
  \documentclass[10pt,preprint,pdftex]{aastex}
\fi
\newcommand{\rmmat}[1]{{\hbox{\rm #1}}}
\newcommand{\rmscr}[1]{{\rmmat{\scriptsize #1}}}
\newcommand{\be}{\begin{equation}}
\newcommand{\ee}{\end{equation}}
\newcommand{\eq}[1]{\be{#1}\ee}
\newcommand{\bt}{\begin{table} \begin{center}}
\newcommand{\et}{\end{center} \end{table}}
\newcommand{\ba}{\begin{eqnarray}}
\newcommand{\ea}{\end{eqnarray}}

\newcommand{\dd}[2]{\frac{d #1}{d #2}}

\newcommand{\eqref}[1]{Equation~(\ref{eq:#1})}

\begin{document}
\newcommand{\bfi}{{\bf B}} \newcommand{\efi}{{\bf E}}
\newcommand{\lel}{{\lambda_e^{\!\!\!\!-}}}
\newcommand{\me}{m_e}
\newcommand{\mcs}{{m_e c^2}}
\newcommand{\ho}{{\hat {\bf o}}}
\newcommand{\hm}{{\hat {\bf m}}}
\newcommand{\hx}{{\hat {\bf x}}}
\newcommand{\hy}{{\hat {\bf y}}}
\newcommand{\hz}{{\hat {\bf z}}}
\newcommand{\hr}{{\hat {\bf r}}}
\newcommand{\omv}{\mathbf{\omega}}
\newcommand{\jpa}{Journ. Phys. A}

\title{A QED Model for Non-thermal Emission from SGRs and AXPs}

\author{Jeremy S. Heyl$\footnote{Canada Research Chair}$} 
\affil{
Department of Physics and Astronomy, 
University of British Columbia 
6224 Agricultural Road, Vancouver, British Columbia, Canada, V6T 1Z1
}
\author{Lars Hernquist} \affil{
Harvard-Smithsonian Center for Astrophysics, MS-51, 
60 Garden Street, Cambridge, Massachusetts 02138, United States}

\begin{abstract}

Previously, we showed that, owing to effects arising from quantum
electrodynamics (QED), magnetohydrodynamic fast modes of sufficient
strength will break down to form electron-positron pairs while
traversing the magnetospheres of strongly magnetised neutron stars.
The bulk of the energy of the fast mode fuels the development of an
electron-positron fireball.  However, a small, but potentially
observable, fraction of the energy ($\sim 10^{33}$~ergs) can generate
a non-thermal distribution of electrons and positrons far from the
star.  In this paper, we examine the cooling and radiative output of
these particles.  We also investigate the properties of non-thermal
emission in the absence of a fireball to understand the breakdown of
fast modes that do not yield an optically thick pair plasma.  This
quiescent, non-thermal radiation associated with fast mode breakdown
may account for the recently observed non-thermal emission from
several anomalous X-ray pulsars and soft-gamma repeaters.

\end{abstract}

\section{Introduction}
\label{sec:introduction}

Soft gamma repeaters (SGRs) and anomalous X-ray pulsars (AXPs)
comprise a subclass of neutron stars with unusual properties compared
to typical radio pulsars.  The dozen or so known SGRs and AXPs emit
pulsed X-rays, with steadily increasing periods, as well as bursts of
hard X-rays and soft $\gamma$-rays at irregular intervals, spin
slowly, and appear to be radio quiet.  Compared with high-mass X-ray
binaries, SGRs and AXPs have relatively low X-ray luminosities, soft
spectra, and no detectable companions 
\citep[for reviews, see, e.g.][]{2000AIPC..510..515H,2002nsps.conf...29M}.

While it seems clear that SGRs and AXPs cannot be rotation-powered,
their ultimate energy source remains unclear.  In the ``magnetar''
model \citep{Dunc92,Thom95}, these objects are ultramagnetised neutron
stars with fields $B_* \sim 10^{14} - 10^{15}$ G, fueled by either
magnetic field decay \citep{Thom96,Heyl98decay} or residual thermal
energy \citep{Heyl97kes,Heyl97magnetar}.  Alternatively, these stars
may have ``ordinary'' field strengths $B_* \sim 10^{11} - 10^{13}$ G
and be powered by accretion from e.g. a fossil disk
\citep{1995ApJ...443..786C,2000ApJ...534..373C,2001ApJ...554.1245A,
menou01},
but recent analyses indicate that disks of the required mass may not
survive sufficiently long around these stars
\citep[e.g.][]{2005astro.ph..1551E}.

 According to the magnetar model, bursts occur when fractures in the
 crust of the star send Alfv\'en waves into the magnetosphere
 \citep{Thom95}.  In earlier work, we analysed wave propagation
 through fields exceeding the quantum critical value $B_\rmscr{QED}
 \equiv m^2c^3/e\hbar \approx 4.4\times 10^{13}$ G, and demonstrated
 circumstances under which electromagnetic \citep{Heyl98shocks} and some
 MHD waves, particularly fast modes \citep{Heyl98mhd} evolve in a
 non-linear manner and eventually exhibit discontinuities similar to
 hydrodynamic shocks, owing to vacuum polarisation from quantum
 electrodynamics (QED).  In \citet[hereafter Paper I]{Heyl03sgr}, we
 developed a theory to account for bursts from SGRs and AXPs based on
 ``fast-mode breakdown,'' in which wave energy is dissipated into
 electron-positron pairs when the scale of these discontinuities
 becomes comparable to an electron Compton wavelength.  We showed
 that, under appropriate conditions, an extended, optically thick
 pair-plasma fireball would result, radiating primarily in hard X-rays
 and soft $\gamma$-rays.

Our mechanism provides a natural and efficient outlet for
ultramagnetised neutron stars to form bursts of high energy radiation
without significant emission at shorter wavelengths.  However, as we
noted in Paper I, non-thermal electrons and positrons would be
produced far from the star, beyond the optically thick fireball,
radiating over a wide range of frequencies.  In addition, MHD fast
modes of insufficient amplitude to generate an optically
thick fireball will still dissipate through pair-production, seeding
non-thermal emission.

In what follows, we extend our previous investigation of fast-mode
breakdown to estimate the spectrum of non-thermal emission expected
outside the region containing an optically thick fireball.  We also
consider the fate of fast modes which dissipate their energy through
pair production but at a rate insufficient to yield a fireball.
We show that when a fireball is generated with sufficient energy to
account for the thermal radiation associated with bursts from SGRs
or AXPs, small, but detectable amounts of non-thermal emission
are predicted which dominate the spectrum both well below
and well above X-ray energies.  Our model makes a specific prediction
for the spectral energy distribution of the non-thermal component
that can be used to test our theory for SGR and AXP bursts.

For weaker fast modes that do not yield an optically thick fireball,
we show that the non-thermal radiation may be sufficient to explain
the quiescent, non-thermal emission observed recently from several
SGRs and AXPs.  Therefore, our model naturally provides a unified
explanation for several of the unique characteristics of SGRs and
AXPs in the context of the magnetar scenario.

\section{Calculations}
\label{sec:calculations}

In Paper I, we assumed that the energy dissipated from the fast mode
fuels a pair fireball.  This assumption is reasonable as long as the
region is opaque to X-rays.  Any electrons and positrons produced will
generate synchrotron radiation which will be scattered and form
additional pairs in the strong magnetic field.  Beyond about one
hundred stellar radii, the synchrotron radiation can escape without
becoming thermalised.  Although it may suffer synchrotron
self-absorption, a power-law distribution of electrons and positrons
can radiate outside the fireball, and we would expect a synchrotron
radiation spectrum.

\subsection{Primary Pairs and Photons}
\label{sec:prim-pairs-phot}

Below, we examine our model from Paper I in the limit when the pair
plasma does not thermalise.  As fast modes produced near the stellar
surface propagate outwards, they develop strong discontinuities
(shocks) when the magnetic field is sufficiently large and the wave
energy will eventually be dissipated through pair production.  If we
consider the situation when the electron-positron distribution is no
longer thermal, we must adopt some model for the energies of these
particles when they are created.  A simple and reasonable assumption
is that they are nearly at rest in the frame of the shock as they are
produced.  In the ultrarelativistic limit, this is required in order
to conserve both momentum and energy as the wave dissipates.

The shock is ultrarelativistic and travels at the speed of light in
the strong magnetic field.  The non-thermal pairs are produced far
from the star where it is appropriate to use the weak-field limit of the
index of refraction,
\eq{
n_\perp \approx 1 + \frac{\alpha_f}{4\pi} \frac{8}{45} \left (
\frac{B}{B_\rmscr{QED}}  \right )^2  \, ,
\label{eq:1}
}
where $\alpha_f \equiv e^2 / c \hbar$ is the fine-structure constant, and
so
\eq{
\gamma = \frac{1}{\sqrt{1-\beta^2}} \approx
\frac{1}{\sqrt{2(1-\beta)}} \approx \frac{3}{2}
\sqrt{\frac{5\pi}{\alpha_f}} 
\frac{B_\rmscr{QED}}{B}
\label{eq:2}
}
where $B_\rmscr{QED} \equiv m^2 c^3 / (e \hbar) \approx 4.4 \times 10^{13}$~G.
(We note that the derivation of the effective Lagrangian of QED
presented by \citet{Heyl97hesplit},
valid for any magnetic field strength, correctly reduces to the
weak field limit of e.g. \citet{Heis36},
yielding the above expression for the index of refraction
\citep{Heyl97index}.)

Paper I describes in detail how the energy is dissipated as the fast
mode travels away from the stellar surface.  Here, we are interested in
the evolution of the wave in the weak-field limit, far from
the star, so we can obtain
simple relationships for the amount of energy released as a function
of the distance from the surface.  The emission from outside a
particular radius is proportional to $(k b)^{-2} r^{-3}$ where $k$ is
the initial wavenumber of a fast-mode wave, $b$ is its initial amplitude
compared to the magnetic field of the star, and $r$ is the distance
from the star, so we have
\eq{
d E = A r^{-4} dr \, ,
\label{eq:3}}
where the constant $A$ must be determined from modeling the
evolution of the wave near the neutron star, as described in Paper I.

At each radius from the star, we assume that the pair spectrum is
monoenergetic, so the total emission is the sum of contributions from
different electron-positron energies traveling in different strengths
of magnetic field.  The particles more distant from the star have
higher typically energies and also travel through weaker fields.

We explicitly add the radial dependence to the Lorentz factor of
the particles and calculate the Larmor radius of the particles as a
function of distance from the star.
\eq{
\gamma \approx 70 \frac{B_\rmscr{QED}}{B_*} \left ( \frac{r}{R} \right )^3\, ,
\label{eq:4}}
where $R$ is the radius of the star, and $B_*$ is the pole field strength
at the stellar surface, so
\eq{
r_\rmscr{Larmor} \approx \frac{\gamma m c^2}{e B} = \frac{3}{2}
\sqrt{\frac{5\pi}{\alpha_f}} \left ( \frac{B}{B_\rmscr{QED}} \right
  )^{-2} \frac{\hbar}{m c} = 2.6 \times 10^{-15} r R_6^{-1}
\left (\frac{r}{R}\right)^5
\left ( \frac{B_*}{B_\rmscr{QED}} \right   )^{-2} \, ,
\label{eq:5}}
where $R_6$ is the radius of the star in units of $10^6$ cm,
and we see that particles produced closer than a thousand stellar radii
are essentially trapped.  From Paper I, we find that the fireball
extends to about one hundred stellar radii, so our region of interest
spans an order of magnitude in distance but three orders of magnitude
in magnetic field strength and particle energy.

If classical synchrotron radiation is important, the emission at each 
radius will extend approximately up to photon energies of
\eq{
\hbar \omega =  \gamma^2 \hbar \omega_B =  \gamma^2
\frac{e B}{m c} \approx \frac{45}{4} \frac{\pi}{\alpha_f}
\frac{B_\rmscr{QED}}{B} m c^2 \, ,
\label{eq:6}}
where $\hbar\omega_B$ is the non-relativistic cyclotron energy.

We can estimate the recoil on the electron owing to the emission
of individual photons
\eq{
 \chi = \frac{\hbar \omega}{\gamma mc^2} =  \gamma
\frac{B}{B_\rmscr{QED}} = \frac{3}{2} \sqrt{\frac{5\pi}{\alpha_f}}
\approx 70 \gg 1 \, ,
\label{eq:7}
}
and, so, we must treat the synchrotron emission quantum mechanically.

According to the results of \citet{Land4}, the emission peaks at a
energy
\eq{
E_\rmscr{peak} = \frac{1}{2} \hbar \omega_c = \frac{1}{2} \gamma m c^2\frac{
  \chi}{\frac{2}{3} + \chi} \approx \frac{1}{2} \gamma m c^2 = 
\frac{3}{4} \sqrt{\frac{5\pi}{\alpha_f}} \frac{B_\rmscr{QED}}{B} m c^2 \, .
\label{eq:8}}
The power radiated by the electron is given by
\eq{
I \approx \frac{32 \Gamma \left ( \frac{2}{3} \right )}{243} \alpha_f
 \left (3 \chi \right )^{2/3} \frac{(m c^2)^2}{\hbar} \approx 0.05 \frac{(m c^2)^2}{\hbar} 
\label{eq:9}
}
so the electrons and positrons lose their initial energy over $\sim20
\gamma$ Compton wavelengths and the primary radiation is strongly
beamed parallel to the direction of the fast wave.

\subsection{Formation of Secondary Particles}
\label{sec:form-second-part}

The high-energy synchrotron photons passing through the strong
magnetic field may produce pairs through one-photon pair-production.
It is reasonable to assume that $\hbar \omega \gg\ mc^2$
(Eq.~\ref{eq:8}), so we will use the results of \citet{Land4}.  The
key parameter is essentially
\eq{
\kappa = \frac{\hbar^2 e B \omega}{m^3 c^5} = \frac{B}{B_\rmscr{QED}}
\frac{\hbar \omega}{m c^2} \approx \frac{\chi}{2} = \frac{3}{4} \sqrt{\frac{5\pi}{\alpha_f}}
\approx 35 \gg 1 \, ,
\label{eq:10}
}
where we have used the value of $\omega_c$ from Eq.~(\ref{eq:6}).
Therefore, we can use the large $\kappa$ limit of the pair-production
rate 
\eq{
w =\frac{ 5 \Gamma^2 \left ( \frac{2}{3} \right ) }{7 \pi^{1/2} \Gamma
  \left ( \frac{7}{6} \right ) }
\frac{3^{1/6}}{2^{4/3}} \alpha_f \frac{B}{B_\rmscr{QED}}
\frac{mc^2}{\hbar} \kappa^{-1/3} \approx 8.5 \times 10^{-4}
\frac{mc^2}{\hbar} \frac{B}{B_\rmscr{QED}} \, .
\label{eq:11}
}
These secondary pairs will typically have energies one-quarter that
of the first generation of particles.  The fourth generation of
particles will have energies typically $4^{-3}$ or 64 times smaller
than the first generation.  This fourth generation will have $\chi
\sim 1$ so the typical energy of the resulting synchrotron photons
will be less than one-half the energy of the pairs so $\kappa  \ll 1$,
quenching the pair production rate,
\eq{
w = \frac{3^{3/2}}{2^{9/2}} \alpha_f \frac{B}{B_\rmscr{QED}}
\frac{mc^2}{\hbar} e^{-8/3 \kappa} \, .
\label{eq:12}
}
We assume that that $c/w \lesssim r$ for pair production to be
effective.  We find that the pair-production quenches for 
\eq{
\kappa \sim 0.1 \left [ 1 + 0.04  \ln \left (
  \frac{r}{10^8~\rmmat{cm}} \frac{B}{10^{-5} B_\rmscr{QED}} \right ) \right ]^{-1} \, .
\label{eq:13}
}
To calculate the final spectrum of radiation, we will make the ansatz
that only the final generation of pairs
contributes to the observed emission and that their energies are
about half the energy of the last generation of pair-producing 
photons, so they have $\chi \approx \kappa/2 \sim 0.04$. 
Using the definition of $\chi$ we find
\eq{
\gamma = \chi \frac{B_\rmscr{QED}}{B} = 0.05 \frac{B_\rmscr{QED}}{B} \, .
\label{eq:14}
}
If we compare this result with Eq.~(\ref{eq:2}), we find that the
energy of the pairs has been degraded by a factor of about 1700.

Because $\chi \ll 1$ we can assume that the emission is classical.
The emission will have typical energies of 
\eq{
\hbar \omega_t =  \gamma^2 \hbar  \omega_B \approx 2.5 \times
10^{-3} \frac{B_\rmscr{QED}}{B} m c^2 \, .
\label{eq:15}}

We can combine Eq.~(\ref{eq:3}) with Eq.~(\ref{eq:15}) to obtain an
estimate of the spectrum of resulting photons,
\eq{
d E \propto r^{-6} d E_\gamma \propto E_\gamma^{-2} d E_\gamma
~\rmmat{for}~E_\gamma > E_\rmscr{break} \, ,
\label{eq:16}
}
where $E_\rmscr{break}$ is the value of Eq.~(\ref{eq:15}) at the edge
of the fireball.

A final important element is the cooling time of the final generation
of pairs:
\eq{
\tau_\rmscr{cool} = \frac{\gamma m c^2}{\frac{2}{3} r_0^2 c
  \beta_\perp^2 \gamma^2 B^2} \approx \frac{30}{\alpha_f}
 \frac{B_\rmscr{QED}}{B} \frac{\hbar}{mc^2} \, ,
\label{eq:17}}
where $r_0 \equiv e^2/mc^2$ is the classical electron radius, and
so the final generation of pairs cools nearly immediately.  The pairs 
closest to the surface of the star have the lowest energy and cool the
most quickly, so for lower energies we have
\eq{
d E \propto E_\gamma^{-1/2} d E_\gamma ~\rmmat{for}~
E_0 < E_\gamma <
E_\rmscr{break} \, ,
\label{eq:18}}
where $E_0=\hbar \omega_{B,\rmscr{FB}}$ is the non-relativistic cyclotron
energy at the outer edge of the fireball.   Below this energy we observe
radiation from the pairs formed further away from the surface of
the star.  There are fewer of these pairs, but because they form in a
weaker magnetic field their spectra will extend to lower energies.

The spectra of these electrons extends as $E^{-1/2}$ from
$E_\rmscr{break} B_\rmscr{FB}/B$ down to $E_0 B/B_\rmscr{FB}$ but the
normalisation is smaller by $(B/B_\rmscr{FB})^2$ so we have 
\eq{ 
d E
\propto E_\gamma d E_\gamma ~\rmmat{for}~ E_\rmscr{min} < E_\gamma <
E_0 \, ,
\label{eq:19}
}
where $E_\rmscr{min}$ is the cyclotron energy at the radius where
the pairs are no longer trapped from Eq.~\ref{eq:5}.

Putting together Eq.~(\ref{eq:16}),~(\ref{eq:18}) and~(\ref{eq:19}) 
and normalising by the 
total energy in non-thermal radiation yields the complete spectrum
\eq{
\dd{E}{E_\gamma} = \frac{E_\rmscr{total}}{3 E_\rmscr{break}}
\left \{
\begin{array}{lc} 
\frac{E_\gamma}{E_\rmscr{break}} \left
(\frac{E_\rmscr{break}}{E_0} \right )^{3/2}
, & E_\rmscr{min} < E_\gamma <E_0  \\
\left ( \frac{E_\gamma}{E_\rmscr{break}}\right )^{-1/2}, & E_0
< E_\gamma <
  E_\rmscr{break} \\
\left ( \frac{E_\gamma}{E_\rmscr{break}} \right )^{-2}, &
E_\rmscr{break}  < E_\gamma < E_\rmscr{max}
\end{array}
\right .
\label{eq:20}
}
where $E_\rmscr{min}$ and $E_\rmscr{max}$ are determined by the
non-relativistic cyclotron frequency and the typical emission energy,
given in Eq.~\ref{eq:16} at the outer edge of the region where the
primary pairs are initially trapped by the field from Eq.~\ref{eq:5}.  
Typically, $E_\rmscr{min} \sim 10^{-3} E_0$ and
$E_\rmscr{max} \sim 10^3 E_\rmscr{break}$.  To find the normalisation 
in Eq.~(\ref{eq:20}), we have assumed that $E_\rmscr{max} \gg
E_\rmscr{break}$ and $E_0 \ll E_\rmscr{break}$

The bulk of the energy emerges near $E_\rmscr{break}$, and the bulk of
the photons have energies near $E_0$, and the
mean energy of a photon is 
\eq{
\sqrt{E_\rmscr{break} E_0} = \chi mc^2 = 0.05 mc^2 \, ,
\label{eq:21}
}
where $\chi$ is given in Eq.~(\ref{eq:14}).

We now estimate the values of $E_0$ and $E_\rmscr{break}$.  From the
models presented in Paper I, we have $r_\rmscr{FB} \approx 100 R$, so 
\eq{
E_0 \approx 10^{-6} \frac{B_*}{B_\rmscr{QED}} mc^2 \approx
5~\rmmat{eV}~\rmmat{and}~E_\rmscr{break} \approx 1600
\frac{B_\rmscr{QED}}{B_*} mc^2
\approx 80~\rmmat{MeV} \, ,
\label{eq:22}
}
where the final results assume that $B_*=10B_\rmscr{QED}$.  The mean
photon energy is around 25~keV.  In Paper I, we argued that about
$10^{34}$~ergs per burst may go into non-thermal emission.  In this
case, we would have approximately $2.5\times 10^{41}$ photons or
$2 \times 10^{-3}$ photons per square centimeter at a distance of
10~kpc.  The bulk of the photons will have energies around $E_0$.
\begin{figure}
\plotone{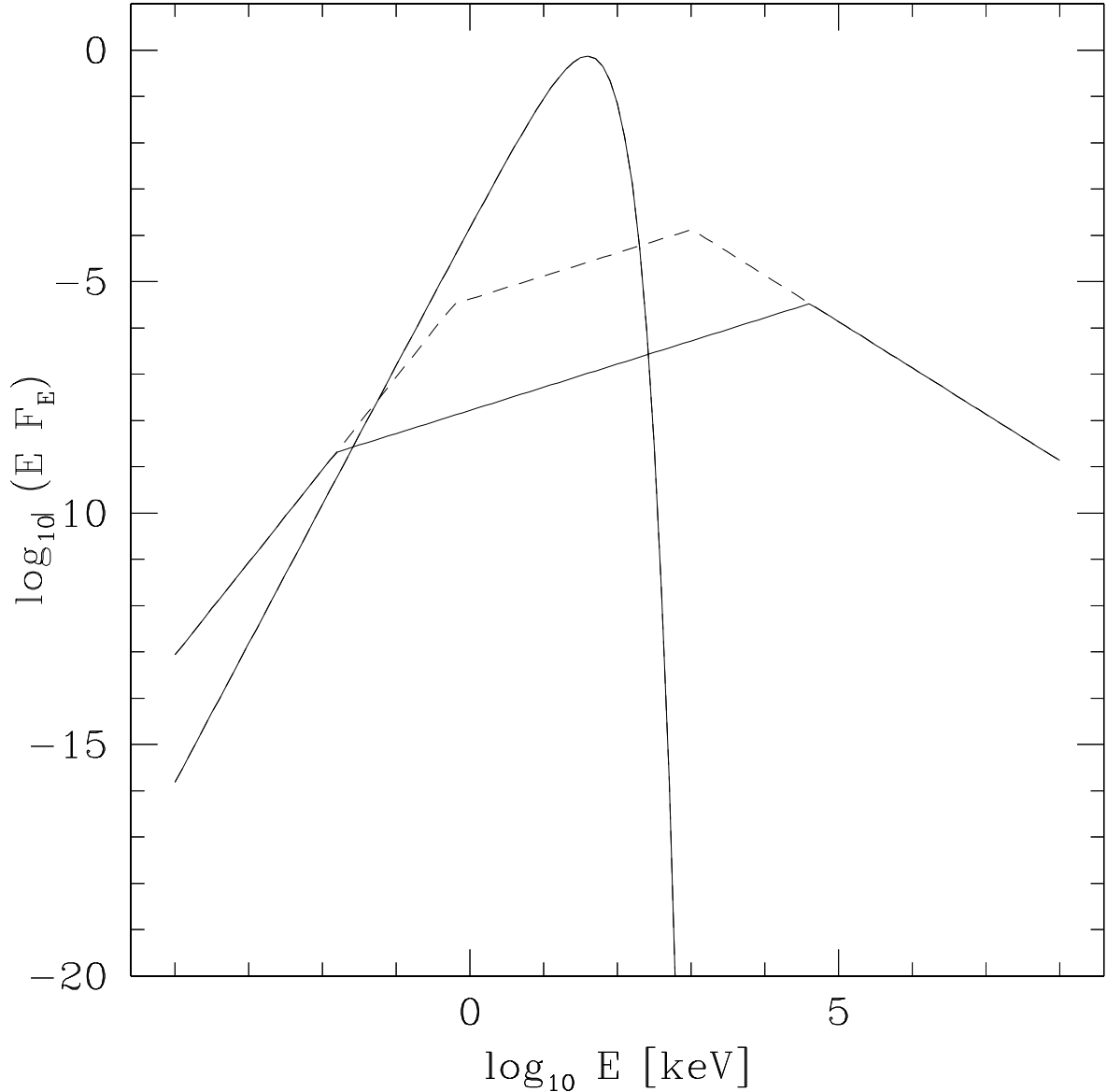}
\caption{Spectrum from an SGR 
  or AXP fireball (smooth solid curve) compared with the non-thermal
  emission.  $T_{FB}=10$~keV and the non-thermal emission comprises
  $10^{-5}$ of the fireball energy.  The value of the magnetic field
  at the edge of the fireball (100 stellar radii) is taken to be $3
  \times 10^{-5} B_\rmscr{QED}$, yielding $E_0=15$~eV and
  $E_\rmscr{break}=40$~MeV (broken solid line).  
   If the fireball ends at thirty stellar
  radii, $E_0$ and $E_\rmscr{break}$ increase and decrease
  respectively by a factor of 25000 
  and the total non-thermal energy is $3\times 10^{-4}$ of
  the fireball emission (dashed line).  As we argue in 
  \S~\ref{sec:fast-mode-cascade}, values of $E_\rmscr{break}$ less than
  1~MeV do not make sense physically, so the fast-mode cascade as we have
  described is limited   to fields less than $\approx
  10^{-3}B_\rmscr{QED}$, well within the weak-field limit.
  }
\label{fig:spec}
\end{figure}

The total energy in the non-thermal emission is proportional to the
value of the magnetic field at the edge of the fireball (from
Eq.~\ref{eq:3}), so as one can see from Fig.~\ref{fig:spec}, the
high-energy spectrum simply extends to a lower value of
$E_\rmscr{break}$ and the low-energy spectrum extends to a higher value
of $E_0$ as the radius of the fireball decreases.

\subsection{Time Dependence of Spectral Emission}
\label{sec:time-depend-spectr}

Fig.~\ref{fig:spec} gives the total spectrum of radiation, both thermal
and non-thermal from the burst, integrated over time.  While the pairs
are still relativistic we find that the cooling time is
\eq{
\tau_\rmscr{cool} = \frac{3}{2} \left ( \frac{E_t}{mc^2}
\frac{B}{B_\rmscr{QED}} \right )^{-1/2} \frac{\hbar}{mc^2}
\label{eq:23}
}
as a function of the energy of the typical synchrotron photon
$E_t$.  If we compare this result with Eq.~(\ref{eq:15}), we find that
initially the typical photon energy is also inversely proportional to
the field strength, so as time passes we have a spectrum 
$F_E \propto E_\gamma^{-2}$ above a frequency 
$E_\rmscr{break}(t) \propto t^{-2}$.  The value of $F_E$ at
$E_\rmscr{break}(t)$ is independent of time.  

The pairs produced nearest to the star become non-relativistic first
so as $E_\rmscr{break}$ approaches $E_0$ the spectrum steepens and at
late times approaches $F_E \propto E_\gamma$, but now the normalisation 
decreases rapidly with time.   To appreciate the rate of cooling we
can substitute a value for $E_0$ into the cooling time Eq.~(\ref{eq:23})
to get
\eq{
  \tau_\rmscr{cool} = \frac{3}{2} 
  \frac{B_\rmscr{QED}}{B} \frac{\hbar}{mc^2}
}
so the pairs cool very rapidly to non-relativistic energies over
$10^{5-8}$ Compton times or 0.1 to 10 femto-seconds.  Observationally,
one expects to see the non-thermal spectrum as depicted in
Fig.~\ref{fig:spec} immediately before the fireball radiation ($\sim
10^{-3}$~s).  The non-thermal emission dominates the fireball emission
both at low energies and for $E_\gamma \gg\ kT_{FB}$

\subsection{Inverse-Compton versus Synchrotron Power}
\label{sec:inverse-compt-vers}

We have assumed  implicitly that synchrotron emission dominates the
energy loss of the high energy pairs produced in the cascade.  If we
assume that the fireball emits $10^{40}$~erg/s we can estimate the
photon energy density:
\eq{
U_\rmscr{photon} = \frac{L}{4\pi r^2 c} = 2 \times 10^{12} L_{40}
r_8^{-2}~\rmmat{erg cm}^{-3}
}
and the magnetic energy density
\eq{
U_B = \frac{B^2}{8\pi} = 4 \times 10^{16} B_{*,15}^2 r_8^{-6} ~\rmmat{erg cm}^{-3}
}
so
\eq{
\frac{P_\rmscr{IC}}{P_\rmscr{Sych}} = \frac{U_\rmscr{photon}}{U_B} =
5\times 10^{-6} L_{40} r_8^4 B_{*,15}^{-2} \, .
}
Inverse-Compton emission may be important for $r \gg 10^9$~cm.
Coincidentally, this is the same distance from the star where the
Larmor radius of the initial pairs becomes greater than the distance
from the star (Eq.~\ref{eq:5}), where synchrotron emission is not very
effective.  Regardless, the total energy deposited in 
this outermost region is $10^{-3}$ lower than the synchrotron-dominated
region (Eq.~\ref{eq:3}).

In principle, the larger bursts such as the March 7 and August 28
events have sufficient luminosity to make inverse Compton emission
important.  However, the cooling time for the pairs (Eq.~\ref{eq:17})
is so short that the pairs will lose most of their initial energy before the
fireball radiation reaches the non-thermal emitting region.  The
cooling time for pairs emitting at lower energies is sufficiently long
that inverse-Compton emission might be important for these electrons
in the brightest SGR bursts.

\subsection{Two-photon Pair Production}
\label{sec:two-photon-pair}

To determine the efficacy of two-photon pair production we must first
decide whether the highest energy photons produced in the
cascade (Eq.~\ref{eq:8}) lie above the threshold to 
interact with the thermal photons from the neutron star or other
ambient photons.   The photons produce by the fireball simply do not
catch up with the photons produced in the cascade.

For the thermal photons, we must have
\be
E_\rmscr{CM} = \sqrt{E_\rmscr{peak} E_\rmscr{thermal} \left ( 1 -
  \cos\theta \right )} \approx 0.133 \left (
\frac{E_\rmscr{thermal}}{1~\rmmat{keV}} \frac{B_\rmscr{QED}}{B_*}
  \frac{r}{R} \right)^{1/2}~\rmmat{MeV} > 1~\rmmat{MeV}
\ee
so for
\be
r > 60 \frac{B_*}{B_\rmscr{QED}} \left (
\frac{E_\rmscr{thermal}}{1~\rmmat{keV}} \right)^{-1}
R
\ee
two-photon pair production may operate.  However, the cascade
typically dumps most of its energy around $r=100 R$ and $B_* \gtrsim
10 B_\rmscr{QED}$ so two-photon pair production against the thermal
photons from the neutron star lies below threshold.   Furthermore,
even if the thermal photons were above threshold, the mean-free path is
greater than $10^3$~R and much much greater than the mean-free path
for one-photon pair production.

We can repeat the calculation for ambient photons from starlight
\be
E_\rmscr{CM} = \sqrt{ E_\rmscr{peak} E_\rmscr{ambient} } = 4\times
10^{-3} \left ( \frac{ E_\rmscr{ambient}}{1~\rmmat{eV}}
\frac{B_\rmscr{QED}}{B_*} \frac{r^3}{R^3} \right )^{1/2}~\rmmat{MeV}
\ee
so
\be
r > 40 \left ( \frac{B_*}{B_\rmscr{QED}} \right )^{1/3} \left
(\frac{E_\rmscr{ambient}}{1~\rmmat{eV}}\right)^{-1/3} R.
\ee
The ambient starlight typically does have sufficient energy to
pair-produce against the first generation of photons produced in the
cascade but the mean-free path is from kiloparsecs to megaparsecs much
greater than the scale of the cascade region.

\subsection{The Fast-Mode Cascade Contrasted with the Pulsar Cascade}
\label{sec:fast-mode-cascade-1}

The main difference between the fast-mode cascade and the
pair-cascade invoked to explain pulsar emission is geometry.  The
pulsar cascade is instigated by charged particles that are restricted
to travel along magnetic field lines -- any perpendicular momentum is
quickly reduced through synchrotron emission.  The photons that are
produced also travel nearly parallel to the field lines, so photons
must travel some fraction of the curvature length of the magnetic
field before their momentum perpendicular to the magnetic field is
sufficient to produce a pair.  Consequently, the pairs are produced
nearly on threshold.

The fast-mode is not restricted to travel along the field lines.  Here,
we have explicitly assumed that it travels perpendicular to the field
lines.  Regardless of the precise angle, the pairs produced initially
will have much of their momentum perpendicular to the field
resulting in many photons traveling perpendicular to the field that
can in principle produce pairs {\em in situ}.  What quenches the
process is the reduction of the cross-section as the energy of the
subsequent photon generations decreases.

\subsection{The Fast-Mode Cascade without SGR Bursts}
\label{sec:fast-mode-cascade}

In Paper I, we argued that a sufficiently widespread production of
fast modes would result in the creation of a fireball because the
vicinity of the star would be opaque to X-rays simply owing to the
Thomson cross section of the pairs.  The optical depth between the
surface of the star and infinity is simply proportional to the
fraction of the star's surface area generating magnetohydrodynamic
waves.  The SGR bursts typically involve a large surface area $\sim
R^2$ emitting fast-modes with a fractional amplitude of several
percent within several light-crossing times of the inner
magnetosphere.

These large-scale events are bound to be rare.  On the other hand,
smaller scale stresses in the crust of the star may be relieved more
regularly.  This would excite only a small fraction of the neutron star
surface, but in principle the amplitude of the waves might be
sufficient to cause fast-mode breakdown.  In this case, all of
the radiation produced would follow the non-thermal distribution
outlined in this paper.  

What would be the typical flux of these small-scale fast modes?
\citet{Thom96} and \citet{Heyl98decay} have argued that the quiescent
emission of SGR and AXP neutron stars may be powered by the
decay of the magnetic field.  The quiescent thermal emission may only
be a small fraction of the total energy released by the decay of the
magnetic field.  Recent observations of SGRs and AXPs indicate that
the thermal radiation may indeed be just the tip of the iceberg
\citep{2004astro.ph.11696M,2004astro.ph.11695M,2004ApJ...613.1173K};
therefore, we will be liberal in our assumption of the relative amplitude of
the thermal and non-thermal radiation and allow the pair cascade to
operate beyond a certain radius from the star or equivalently below a
certain magnetic field strength ($B_\rmscr{max}$) .   This model has two
parameters, $B_\rmscr{max}$ determines the position of the two breaks
in the spectrum and the total normalisation.  The process of fast-mode
breakdown predicts a particular relationship between the location of
the two breaks in the spectrum~(\ref{eq:22}) and particular
slopes~(\ref{eq:20}) between and beyond the breaks.

\begin{figure}
\plotone{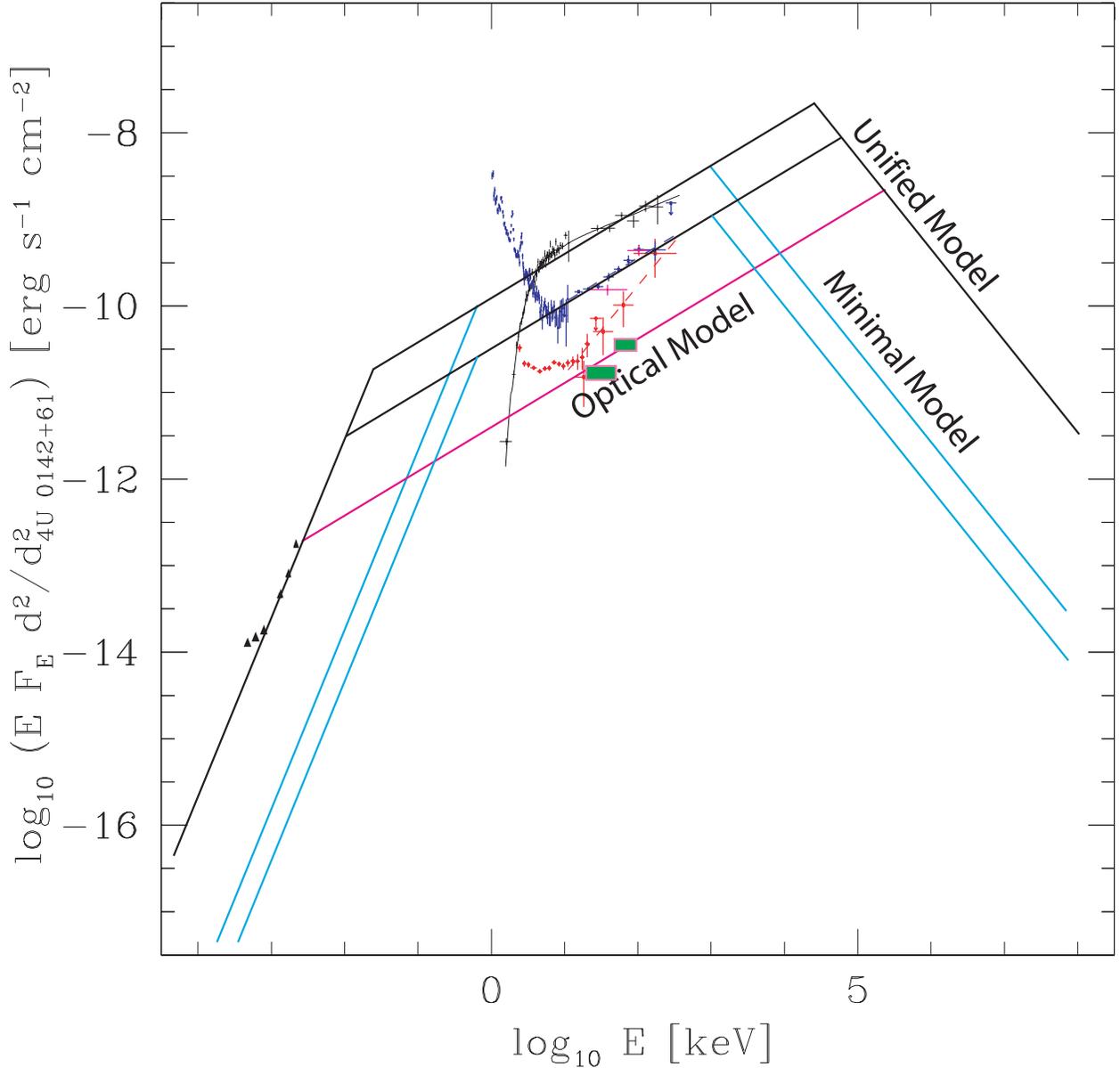}
\caption{The spectrum produced by fast-mode breakdown is superimposed
over the observed thermal and non-thermal emission from several AXPs
and SGRs for models that fit either the optical or INTEGRAL data
solely and one that fits both sets of data.  The unabsorbed optical
data are from \citet{2000Natur.408..689H} via
\citet{2004astro.ph..4144O} for AXP 4U~0142+61.  The uppermost black
symbols are the hard X-ray band are from \citet{2004astro.ph.11696M}
for SGR~1806-20.  \citet{2004astro.ph.11695M} obtained similar results
for the SGR.  The middle sets of points in the hard X-ray data (blue
is total flux and red is pulsed flux) are from
\cite{2004ApJ...613.1173K} for AXP 1E~1841-045.  The green squares
plot the INTEGRAL data reported by \citet{2004ATel..293....1D} for
AXP~4U~0142+61. We normalised the \citet{2004ATel..293....1D} results
using the observations of the Crab by \cite{1989ApJ...338..972J}.  We
scaled the emission from the three sources by assuming that they all
lie at the distance of AXP~4U~0142+61.  We used 3 kpc for
AXP~4U~0142+61 \citep{2000Natur.408..689H}, 7.5 kpc for AXP
1E~1841-045 \citep{1992AJ....104.2189S} and 15 kpc for SGR~1806-20
\citep{2004astro.ph.11696M}.}
\label{fig:allspec}
\end{figure}

Figure~\ref{fig:allspec} shows the observed broad-band spectrum of
several AXPs and SGRs.  To compare the spectra we have placed the more
distant objects (AXP~1E~1841-045 and SGR~1806-20) whose hard X-ray
emission was discovered with INTEGRAL at the distance of 4U~0142+61
whose optical emission \citep{2000Natur.408..689H} is very likely to
be nonthermal \citep{2004astro.ph..4144O}.

Because the location of the two breaks in the spectrum both depend on
the strength of the magnetic field at the inner edge of the breakdown
region, we find that the presence of extensive nonthermal optical
emission indicates that the non-thermal hard X-ray emission should
peak at about 30~MeV, a factor of two hundred beyond the observed
spectrum.  The best limits in this energy range are provided by EGRET.

We come to this conclusion under the assumption that SGR~1806-20 and
AXP~1E~1841-045 have a similar optical excess to 4U~0141+61.  A more
conservative assumption would be that the hard X-ray emission does not
extend far beyond the observations from INTEGRAL with spectral breaks
at about 1~MeV and 650~eV.  This situation is somewhat natural.  The
fast-mode cascade is limited to pairs with sufficient energy to produce
photons with $E > 1$~MeV that can subsequently pair produce.  Lower
energy electrons simply cool, giving the observed cooling spectrum in
the hard X-rays.  The total energy in the non-thermal emission is also
reduced by a factor of a few.  In the context of the fast-mode cascade
it is difficult to have $E_\rmscr{break} < 2 m c^2$.  This model is
denoted as the ``Minimal Model'' in Fig.~\ref{fig:allspec} because
$E_\rmscr{break}$ takes on the minimal value that makes 
sense physically;
i.e $\approx 1$~MeV.
\begin{table}
\caption{Predicted flux above 100~MeV and observed EGRET upper limits \citep{Gren05}  in units of $10^{-8}$ photons s$^{-1}$cm$^{-2}$. The GLAST upper
  limits are nominally $(0.2~\rmmat{--}~ 0.4) \times 10^{-8}$  photons
  s$^{-1}$cm$^{-2}$ \citep{glast}.}
\label{tab:egret}
\begin{center}
\begin{tabular}{l|ccccc}
Object & EGRET            & EGRET       & Unified & Minimal & Optical \\
       & Exposure [weeks] & Upper Limit & Model   & Model   & Model 
 \\ \hline 
AXP~4U~0142+61  & 8.8     &   50        & 1500    & --- & 800 \\
AXP~1E~1841-045 & 6.8     &   70        &   70    & 0.4 & --- \\
SGR~1806-20     & 4.9     &   70        &  280    & 0.6 & --- \\
\end{tabular}
\end{center}
\end{table}

We can also obtain a conservative picture for AXP~4U~0142+61 by
accounting for only the optical excess and getting a prediction for
the hard X-ray emission.  We assume that $E_0$ is a few electron
volts giving $E_\rmscr{break}$ on the order of 200~MeV.  The minimal
hard X-ray emission required to account for the optical emission is
depicted in Figure~\ref{fig:allspec} as the ``Optical Model''.  This
optical model agrees well with the results reported by
\citet{2004ATel..293....1D}.   On the other hand, if AXP~4U~0142+61 has
inherent non-thermal emission as strong as either AXP~1E~1841-045 or
SGR~1806-20, the EGRET flux would be larger as given under the
``Unified Model'' in Table~\ref{tab:egret}.

Table~\ref{tab:egret} gives the predicted EGRET fluxes above 100~MeV
for the various objects and models.  The optical model and the unified
model predict approximately similar EGRET fluxes for 4U~0142+61.
The minimal model based solely on the INTEGRAL data exhibits a flux
above 100~MeV about two hundred times smaller than the unified model.
Because the optical model cannot explain the observed INTEGRAL data
for 1E~1841-045 and SGR~1806-20, it is omitted.  Similarly, the
minimal model cannot explain the optical data for 4U~0142+61.  We see
that for 1E~1841-045 and SGR~1806-20 the predictions for the minimal
model lie comfortably below the EGRET upper limits.  In the context of
the fast-mode breakdown model, this means that the optical emission
for 1E~1841-045 and SGR~1806-20 is inherently weaker than from
4U~0142+61.

On the other hand, 4U~0142+61 is difficult to explain in the context
of either model because of its large optical flux.  Perhaps 4U~0142+61
was more active during the epoch of the optical observations than
during the EGRET observations.  Some AXPs exhibit variable X-ray
emission such as AX J1845-0258 and 1E~1048.1-5937
\citep{2000ApJ...542L..49V,2004ApJ...608..427M} so this conclusion
might be natural.

\section{Discussion}
\label{sec:discussion}

We have presented a model for the non-thermal emission from AXPs and
SGRs.  It is a natural extension of the model for bursts that we
presented in Paper I.  When a dislocation of the surface of the
neutron star is sufficiently large, the resulting fast modes will
produce sufficient pairs to make the inner magnetosphere of the
neutron star opaque to x-rays, generating a fireball.   Some small but
observable fraction of the energy initially in the fast modes is
dissipated outside the opaque region yielding a characteristic
fast-mode breakdown spectrum.  A second possibility is that the crust
of the neutron star is constantly shiftly over small scales,
generating fast modes whose breakdown is insufficient to produce a
fireball.  In this case we would associate the non-thermal radiation 
with the quiescent thermal radiation from the surface of the star.

The exponential quenching of one-photon pair production and the dipole
geometry of the neutron star field at large distances constrains 
the spectrum produced by fast-mode breakdown to have only two free
parameters.  The cyclotron energy ($E_0$) at the inner edge of the
breakdown region (or alternatively $E_\rmscr{break}$) and the
normalisation of the spectrum.  The model does not yield any freedom
in setting the spectral slope.  Below $E_0$, $E F_E$ is proportional to $E$.
Between $E_0$ and $E_\rmscr{break}$ it is proportional to $E^{1/2}$
and above $E_\rmscr{break}$ $E F_E \propto E^{-1}$.

Because of the paucity of free parameters, our model has great
predictive power, if one can ascribe the observed flux from an AXP or
SGR to fast-mode breakdown.  In the case of 4U~0141+61 whose optical
and near infrared flux has been determined
\citep{2000Natur.408..689H}, the hard X-ray flux from fast-mode
breakdown should be at least as large as found by
\citet{2004ATel..293....1D}.  Additionally, we have several
predictions for the gamma-ray flux from AXPs and SGRs.  Regardless of
which model one assumes, the emission should extend beyond observed
INTEGRAL data without a break below 1~MeV.  Further INTEGRAL
observations of these objects may verify this claim.  If the object
has a significant optical excess, such as 4U~0141+61, we predict that
$E F_E$ should continue to rise to $10-200$~MeV, well into the
realm of GLAST.  Even the least favourable ``Minimal'' model for 
AXP~1E~1841-045 and SGR~1806-20 yields a marginal detection with
GLAST.

Observations of the AXPs and SGRs continue to surprise, as do
theoretical investigations of ultramagnetised neutron stars. 
In this paper and Paper I we have presented a unified model for the
thermal burst emission and non-thermal emission from ultramagnetised
neutron stars.  The model has few underlying assumptions: magnetars
produce fast modes sufficient to power the non-thermal emission and,
more rarely, the bursts, the magnetic field far from from the star is
approximately dipolar and quantum electrodynamics can account for the
dynamics of pairs and photons in strong magnetic fields.  The model
for the non-thermal emission has two free parameters, a normalisation
and break frequency.  Further observations can easily verify or
falsify this
model and 
potentially provide direct evidence for the ultramagnetised neutron
stars that power AXPs and SGRs and the macroscopic manifestations of
QED processes that account for their unique attributes.

\acknowledgements
J.S.H. would like to thank Alice Harding for help with the EGRET upper
limits.  J.S.H. is supported by a NSERC discovery grant.

\bibliographystyle{apj}
\bibliography{mine,physics,ns,qed}

\end{document}